\documentclass[cits]{PoS}
\usepackage{natbib}

\title{The peak region of the extragalactic background radiation}

\ShortTitle{The X-ray background peak}

\author{\speaker{Marco Ajello}\thanks{I thank the organizers for the invitation to give this talk and  for organizing a scientifically stimulating event.}\\
 Stanford Linear Accelerator Center, 2575 Sand Hill Road, Menlo Park, CA 94025
and\\
 KIPAC, 2575 Sand Hill Road, Menlo Park, CA 94025\\
        E-mail: \email{majello@slac.stanford.edu}}

\abstract{The Cosmic X-ray background carries the information of cosmic 
accretion onto super-massive black holes. The intensity at its peak 
can be used to constrain the integrated space density of highly obscured AGNs.  Determining the shape and intensity of the Cosmic X-ray background 
radiation represents, however, a first step towards the understanding 
of the population of Compton-thick AGNs.
The study of AGNs in the local and more 
distant Universe allows to understand the whole picture. In this talk, 
I will review 
the current understanding of generation of the Cosmic X-ray background at 
its peak.
I will focus on the recent measurements of the Cosmic X-ray background
 and I will discuss the recent advancements in the understanding of 
AGNs in both the local and more distant Universe. Finally, I will 
also discuss open issues and future prospects. 
}

\FullConference{7th INTEGRAL Workshop\\
		 September 8-11 2008\\
		 Copenhagen, Denmark}

\begin{document}

\section{Introduction}
\label{sec:intro}
\subsection{History of the Cosmic X-ray background}
The discovery of the cosmic X-ray background (CXB) happened
at the same time as the detection of the very first extra-solar
X-ray source \citep{giacconi62}. Both of them  marked the beginning of
a new era in high-energy astrophysics. Since then, the CXB background
has been the object of a lively debate. First about its origin, later
about its spectral shape. Indeed, a diffuse isotropic radiation such
as the CXB might be produced either by hot gas permeating the Universe,
or by millions of point-like X-ray sources or by both. 
Precise measurements done with the HEAO1-A2 experiment \citep{marshall83} 
revealed that the CXB spectral shape is consistent,
between  3--50\,keV with a bremmstrahlung model with a temperature
of 40\,keV. This was seen as a natural evidence for the presence
of a very hot intergalactic medium. This conclusion was supported by,
or was based on, the power-law like emission of Active Galactic Nuclei
\citep[AGNs, e.g.][]{mushotzky84}, whose integrated emission 
(in the case millions of AGNs existed), remains
always a power-law like spectrum. The final resolution of the 
controversy came from the incredibly neat result obtained 
with the FIRAS instrument on board COBE: the absence of any 
detectable deviation from a pure black spectrum body of the cosmic microwave
 background set an upper limit on the contribution of an uniform hot 
intergalactic gas to the CXB of $<10^{-4}$ \citep{wright94}.
Once this issue was solved, it was clear that the CXB emission had to
be the unresolved (at that time) emission of millions of AGNs \citep{setti89}.
The deep X-ray surveys \citep{hasinger93,alexander01,giacconi02} 
have confirmed that indeed a large fraction (80--100\,\%) of the CXB
can be resolved into point-like sources which can then be identified
as AGNs. The discovery that for many AGNs the nuclear radiation is 
partially obscured by intervening matter led population synthesis models
to solve the paradigm of the generation 
of the CXB \citep{comastri95,ueda03,treister05,gilli07} by means of 
AGNs with different amount of absorption.
Although the Chandra and XMM-Newton have resolved most of the Cosmic
X-ray background below $\sim$2\,keV, the fraction of resolved
CXB emission declines with energy being $<50$\,\% above 6\,keV 
\citep{worsley05}. This represents the main evidence for the
presence of a population of AGNs which is still currently undetected
even in the deepest surveys.
The analysis of the  {\it unresolved} component revealed that it might
be consistent with the integrated emission of a population 
of very absorbed, Compton-thick ($\tau=N_H \sigma_T\sim1$ and thus
$N_H\approx1.5\times 10^{24}$\,atoms cm$^{-2}$) AGNs.
Given the fact that their emission is suppressed below 10\,keV,
detecting these object is extremely difficult at soft X-rays and
until a few years ago only a handful of Compton-thick AGNs were known
\citep{comastri04}.

\subsection{Constraints from the Cosmic X-ray background}
Since the direct detection of Compton-thick objects is difficult
in X-rays, the CXB spectrum becomes the final resource to constrain
the space density of such objects. Indeed, the CXB represents the 
integrated emission of the accretion processes onto
 super-massive black holes (SMBHs) present in the Universe.
Integrating the luminosity function of unabsorbed and absorbed
 AGNs \citep{hasinger05,ueda03}, with sensible
hypotheses regarding their spectral properties (and their dispersions),
allows an estimate of the contribution of these two classes to the
total CXB spectrum. The most recent studies \citep{ueda03,treister05,gilli07}
show that the contribution
of these two classes is not enough to explain the totality of the CXB 
highlighting a deficit around the CXB peak at 30\,keV.
Assuming that this deficit emission is due to undetected 
Compton-thick AGNs, it becomes possible to make an estimate of their
space density.  With no other constrains left, except the small number
of heavily obscured objects known in the local Universe, the absolute
normalization of the CXB spectrum (particularly at its peak) represents
the main resource to constrain the Compton-thick population.
This is why it has been much debated lately.
In the 2-10\,keV band the CXB measurements of focusing telescopes
(as XMM-Newton, Chandra, etc.) lie systematically above the one
obtained by non-focusing optics  \citep{revnivtsev05}.
Moreover, it seems that neither cosmic variance 
\citep{barcons00} nor differences in the flux scale calibration of 
each individual instrument \citep{revnivtsev05,frontera07}
may account for this discrepancy. 
This led several authors to naturally question
the broadband measurement of the CXB performed by HEAO1 \citep{gruber99} and 
to use CXB spectra renormalized by a factor $\sim1.3$ 
\citep{ueda03,treister05,worsley05}. A change in the normalization
of the CXB implies either a change in the number density of AGNs
or in their radiative efficiency or in both.  This is why the normalization
of the CXB has been so much debated in the past. In particular the lack
of measurements at the peak of the CXB ($\sim$30\,keV) left the population of 
Compton-thick AGNs loosely constrained. This situation recently improved
with the results shown in the next sections.

\section{Advances in the measurement of the Cosmic X-ray background}
\begin{figure}[ht!]
  \begin{center}
  	 \includegraphics[scale=0.7]{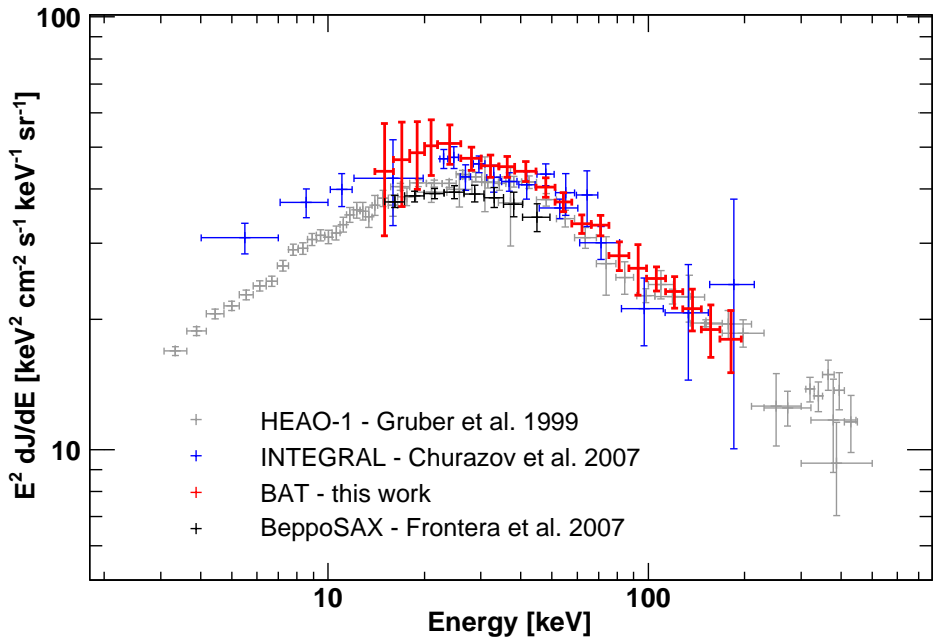}
  \end{center}
  \caption{Comparison of measurements of the peak of the 
Cosmic X-ray background (adapted from \cite{ajello08c}).  
\label{fig:cxb_peak}}
\end{figure}

Measuring the CXB is extremely challenging with currently flying instruments
as none of them has been directly designed to make such measurement. The
main difficulty here is the capability, by design or analysis, to
discriminate the CXB among the other, sometimes dominating, instrumental
background components. In the case of INTEGRAL and {\em Swift}, it is even
worst as their high-energy detectors (IBIS and BAT \citep{ubertini03,barthelmy05}) are coded-mask detectors expressly designed for the study of point-like
sources. Measurements of the CXB have thus to rely on indirect techniques
as the one 
based on the modulation of the CXB signal by the passage of the earth
in the field of view (FOV). All the recent measurements performed
by INTEGRAL, BeppoSAX and {\em Swift}/BAT \citep{churazov07,frontera07,ajello08c}
use the earth occultation technique. These measurements, 
see Fig.~\ref{fig:cxb_peak}, are more consistent
with the original HEAO1 CXB spectrum rather than with its renormalization by 
$\sim1.3$ (see discussion in the previous section). The INTEGRAL and
{\em Swift}/BAT CXB spectra lie $\sim$10\% above the HEAO1 spectrum.
Unfortunately, the albedo emission from the earth, which must be 
taken into account during the occultation analysis, does not allow
to make a strong test of the shape of the CXB spectrum. This test becomes
only possible when different measurements and instruments are used.
Taking into account all the newest measurements and neglecting the broadband
measurement of HEAO1 it is possible to show that a simple description
of the CXB spectrum is achieved using a {\it smoothly joined double
power-law} function \citep{ajello08c}. From this analysis, it appears that the
CXB spectrum, as measured most recently is $\sim$30\,\% larger than the
HEAO1 spectrum below 10\,keV while only 10\,\% larger than it
above this energy.
A new measurement performed in the 1-7\,keV 
band  with the X-ray telescope  (XRT), on board {\em Swift}, confirms
this picture \citep{moretti09}.

\section{Hard X-ray surveys}

\begin{figure*}[ht!]
  \begin{center}
  \begin{tabular}{cc}
    \includegraphics[scale=0.355]{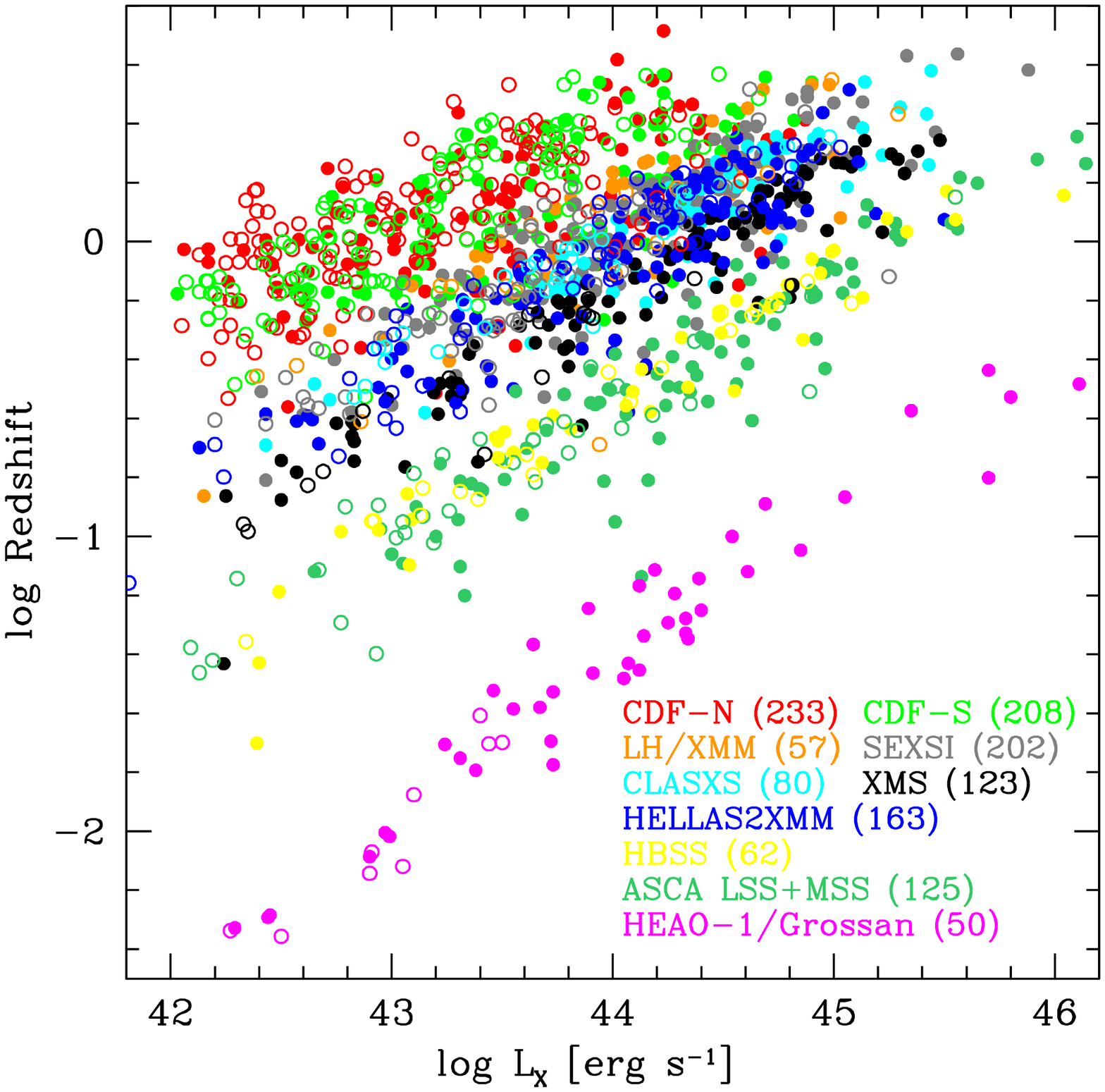} 
  	 \includegraphics[scale=0.40]{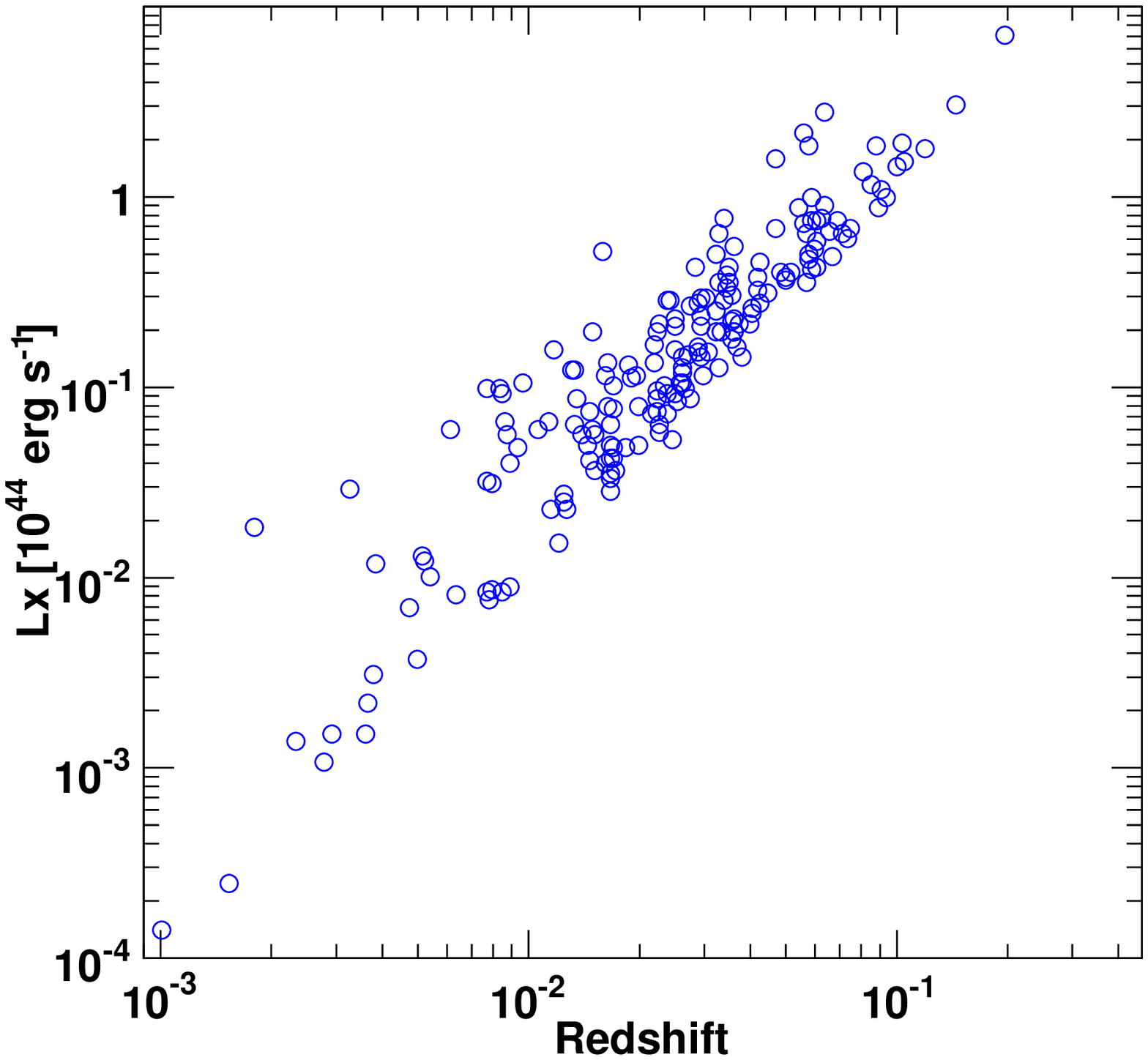}
\end{tabular}
  \end{center}
  \caption{
{\bf Left Panel:} Meta-sample compiled by \cite{hasinger08} using
the available X-ray surveys in the 2-10\,keV energy band.
{\bf Right Panel:} BAT AGN sample (|b|$>$ 15$^{\circ}$) 
from $~\sim$3 years of  all-sky survey in the 15--55\,keV band.
INTEGRAL samples  luminosities and redshifts similar to BAT 
\citep[e.g.][]{beckmann06,bassani06,sazonov07}.
A typical luminosity of $10^{44}$\,erg s$^{-1}$ in the 15-55\,keV 
is equivalent to $\sim 1.2\times 10^{44}$\,erg s$^{-1}$ in the 2-10\,keV
energy band.
 \label{fig:surveys}}
\end{figure*}

\begin{figure}[ht!]
  \begin{center}
  	 \includegraphics[scale=0.7]{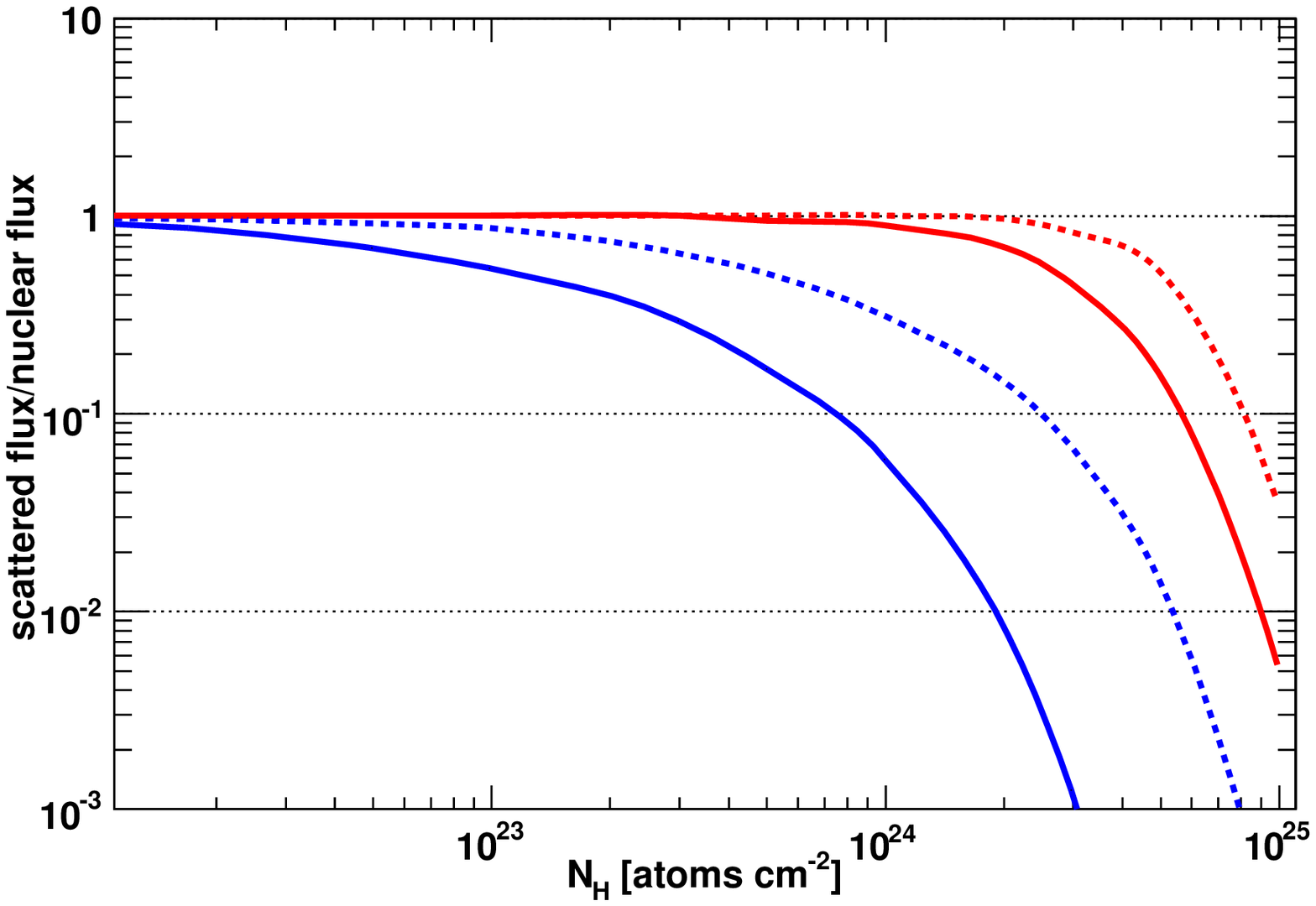}
  \end{center}
  \caption{Ratio of observed flux (scattered) to intrinsic nuclear
flux as a function of column density along the line sight
for the 2--10\,keV (blue) and the 15--55\,keV (red) bands.
The solid and dashed lines are for the case of an AGN at redshift 0 and 
1 respectively.
Compton scattering as well as photoelectric absorption has
been taken into account using the model of \cite{yaqood97}.
The nuclear emission has been modeled as
a power law  with photon index of 2.0.
\label{fig:nh}}
\end{figure}

The understanding of AGN activity, growth and evolution can 
be achieved by combining shallow large area sky surveys  to
very deep pencil beam surveys \citep{hasinger05,lafranca05,silverman08}.
Since most of the AGNs are absorbed (e.g. \citep{gilli07}) 
the best method to select a representative
sample of the entire AGN population is to use an energy band which
is the least affected by attenuation due to intrinsic absorption.
As of now, this is the 2-10\,keV energy band because above this
energy no deep surveys exist. Recently, \cite{hasinger08} compiled
the largest set of X-ray surveys in the 2-10\,keV band.
According to Hasinger 2008, to best determine AGN evolution one should
achieve a good sampling of the luminosity-redshift plane. 
From Fig.~\ref{fig:surveys} (left panel) it is evident that while
Chandra and XMM-Newton are able to perform a good sampling of
the high-luminosity high-redshift part of the plane, the sampling
remain sparse at low luminosities and low  redshifts lacking a large
sample able to constrain the properties of AGNs in the local Universe.
Conversely, the $>10$\,keV range lacks deep surveys 
(able to sample fluxes of $10^{-13}-10^{-14}$\,erg cm$^{-2}$ s$^{-1}$),
but on the other hand, the IBIS and BAT surveys
\citep[e.g.][]{beckmann06,bassani06,sazonov07,ajello08a,tueller08}
are detecting hundreds of AGNs in the local Universe. This is 
clearly shown in the right panel of Fig.~\ref{fig:surveys}.
Moreover, as Fig.~\ref{fig:nh} shows, the $>15$\,keV band
is the best one for an unbiased (with respect to absorption) search of AGNs.
Indeed, for absorbing column densities of 10$^{23}$ and 
10$^{24}$\,atoms cm$^{-2}$ the fraction of nuclear flux which is detected
in the 2--10\,keV band is 50\,\% and 7\,\% respectively.
The 15--55\,keV band, as an example, is unbiased up to 
$\sim10^{24}$\,atoms cm$^{-2}$. This is particularly important
if the goal of the survey is the detection and understanding
of the still {\it missing} Compton-thick source population.

\subsection{The role of absorption}
\label{sec:nh}
The AGN unified models predict that $\sim$3/4 of all AGNs
should be absorbed by intervening
matter along the line of sight. This prediction is based on the 
observations of ionization cones, in O [III] light, with apex
at the nucleus \citep[e.g.][]{evans91}. However, this is clearly
not the case for the X--ray selected AGNs in the local Universe.
Indeed, as Tab.~\ref{tab:nh} shows the fraction of absorbed
AGNs is significantly lower than 75\,\% and closer to $\sim$50\,\%.
Moreover, it is now well established that the fraction of obscured
AGNs is not constant, but that it evolves with luminosity \citep[e.g.][]{lawrence82,ueda03,hasinger08}. This yields that low-luminosity AGNs are surrounded
by an obscuring medium which covers a large solid angle ($80$\,\% of the
sky as seen from the nucleus) while high-luminosity AGNs have a smaller
covering factor and thus must be  able to clean their environment.
 This effect can be interpreted in the framework
of the receding torus model \citep[e.g.][]{lawrence91} as the increase
of the dust sublimation radius as a function of AGN luminosity. This 
would produce a decrease of the solid angle which the dusty torus
covers around the nuclear source, explaining, qualitatively,
the observed trend with luminosity. However, the receding torus
model fails to explain the exact trend of the absorbed fraction with
luminosity \citep[][]{dellaceca08b}. Another option is that 
the torus is not formed by a continuous distribution of dust particles,
but that it comprises of several optically thick clouds orbiting around
the central source \citep[e.g.][]{risaliti07,nenkova08}.
In this framework, the torus consists of a large number of small, self-gravitating, dusty molecular clouds which form a clumpy medium
\citep[e.g.][]{honig07}. This model explains the observed trend
of the absorbed fraction with luminosity very well and this effect might
be  produced
either by a change in the torus width or in the number of clouds or in both
\citep[see e.g.][for details]{nenkova08}.
In any case, the anti-correlation of luminosity and absorption 
has been interpreted as one
of the main evidences for the breakdown of the AGN unified model.

\subsection{Compton-thick AGNs in the local Universe}
\label{sec:cthick_local}
Compton-thick AGNs may come in two flavors: transmission-dominated
and reflection-dominated sources \citep[see e.g.][]{comastri04}.
For trasmission-dominated objects, where a fraction of the nuclear emission 
(piercing through the Compton-thick torus) is detected, 
the spectral shape varies accordingly to the absorbing column density
and the photoelectric cut-off moves progressively to higher energy.
For increasing column densities (24$<$LogN$_H<$25), the absorbing material 
becomes more and more efficient in suppressing the  radiation
below 10\,keV and Compton recoil makes steeper (down-scattering)
the$>$10\,keV part of the spectrum. Transmission-dominated sources
(also called heavily Compton-thick objects) usually show a broad
iron K$_{\alpha}$ line over a flat continuum and the ratio of
the observed to the nuclear flux can be as low as {\it a few} \%
\cite[e.g.][]{comastri07,ueda07}.
Studies of the local Universe in the Optical 
showed that Compton-thick AGNs should
be as numerous as normally obscured ones 
\cite[e.g.][]{risaliti99,guainazzi05}, however up to now only a
handful bona-fide Compton-thick objects were detected 
\citep{comastri04,dellaceca08a}.

Results from the BAT and INTEGRAL surveys allow to shed some light
on the overall picture. Tab.~\ref{tab:nh} shows an up to date
status of hard X-ray surveys. It is apparent that at the fluxes
currently sampled, the fraction of Compton-thick objects is bound
to be $<$20\,\% and likely closer to 10\,\%. Strictly speaking this
represents a lower limit on the real fraction since, as I showed in
Fig.~\ref{fig:nh}, even above $>$15\,keV not all the source flux 
can be detected. This low number of Compton-thick AGNs seems, however,
to be in agreement with the prediction of population synthesis
models which require a substantial contribution from Compton-thick
objects to explain the peak of the CXB emission \cite[e.g.][]{gilli07}.
In other words, Compton-thick AGNs seem to exist, but even 
INTEGRAL and BAT are not sensitive enough to detect many of them
in the local Universe. If these sources evolve similarly
to the other classes of AGNs \cite[e.g.][]{hasinger05,lafranca05},
then their number density is expected to raise quickly with
redshift peaking around $z\approx1$. In this respect, instruments
able to sample the 10$^{-13}$--10$^{-14}$\,erg cm$^{-2}$ s$^{-1}$ fluxes
above 15\,keV
have more chances to detect a large number of heavily absorbed
objects. In this case the {\it k-correction} plays also in their favour
allowing to sample the source spectrum at an higher energy as 
Fig.~\ref{fig:nh} shows.

Another interesting point is that the current hard X-ray surveys 
have not detected any new Compton-thick AGNs of the trasmission-dominated
class. They have found many objects with very high column densities
which are almost Compton-thick (e.g. N$_H\approx10^{23.8-24.0}$\,atoms cm$^{-2}$),
but  not completely so. Quoting \cite{winter08}, who analyzed the 
Tueller et al. (2008) sample,:{\it If we take the Compton-thick 
definition to apply to sources whose column densities 
are $>1.4\times 10^{24}$\,cm$^{-2}$, 
none of the BAT-detected sources are Compton-thick}.
This is also due to the fact that when fitting broad-band X-ray
data (e.g. 0.1-200\,keV), it is difficult (with the low signal-to-noise
ratio at high energy) to discriminate between alternative interpretations.
Indeed, most of the time reflection, partial absorbers and cut-off components
remain degenerate.

\begin{table}[ht]
\centering
\caption{Fraction of absorbed and Compton-thick AGNs, relative to
the whole population for different hard X-ray surveys.
\label{tab:nh} }
\begin{tabular}{lccccc}
\hline
\hline
Reference & Completeness & \% Absorbed   & \% C-thick  & Band & Instrument\\
        & & & & \scriptsize(keV)      \\
\hline
Markwardt et al., 2005   & 95\,\%   & $\sim64$\%  & $\sim$10\% & 15-200 & {\em Swift}/BAT\\
Beckmann et al., 2006    & 100\,\%  & $\sim64$\% & $\sim10\%$ & 20-40 & INTEGRAL\\
Bassani et al., 2006     & 77\,\%   & $\sim65\%$  &   $\sim14\%$ & 20-100 & INTEGRAL\\
Sazonov et al., 2007     & 90\,\%   & $\sim50\%$ & $\sim10$-$15\%$ & 17-60 & INTEGRAL\\
Ajello et al., 2008a     & 100\,\%  & $\sim55\%$  & $<$20\,\% & 14--170 & {\em Swift}/BAT\\
Tueller et al., 2008     & 100\,\%  & $\sim50\%$  &$\sim5\%$ & 14--195 & {\em Swift}/BAT \\
Paltani et al., 2008$^2$     & 100\,\%  & $\sim60\%$ & $<$24\,\%  & 20--60 & INTEGRAL\\
Della Ceca et al., 2008  & 97\,\% & $\sim57\%$   & 0  &4.5--7.5 &XMM-Newton\\
\hline
\end{tabular}
\begin{list}{}{}
\scriptsize
\item[$^{\mathrm{1}}$] The fraction of Compton-thick AGNs comes from 
\cite{sazonov08}.
\item[$^{\mathrm{2}}$] Since the Paltani et al. sample may contain a fraction
of spurious sources, we restricted their sample to a limiting significance of
6\,$\sigma$. Above this threshold all sources are identified 
(see Tab.~2 in \citep{paltani08}).

\end{list}
\end{table}
 
\begin{figure*}[ht!]
  \begin{center}
  \begin{tabular}{cc}
    \includegraphics[scale=0.32]{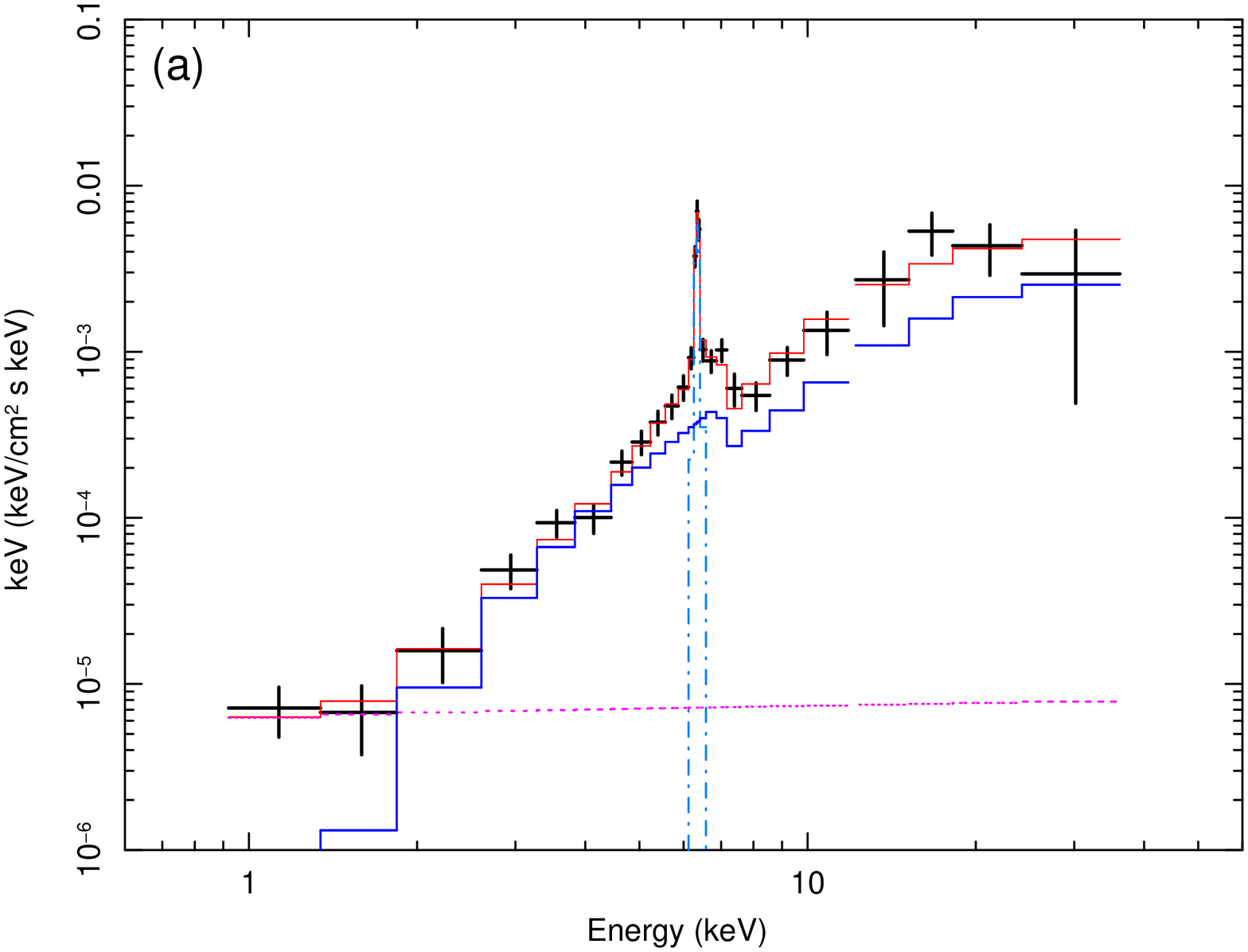} 
  	 \includegraphics[scale=0.30]{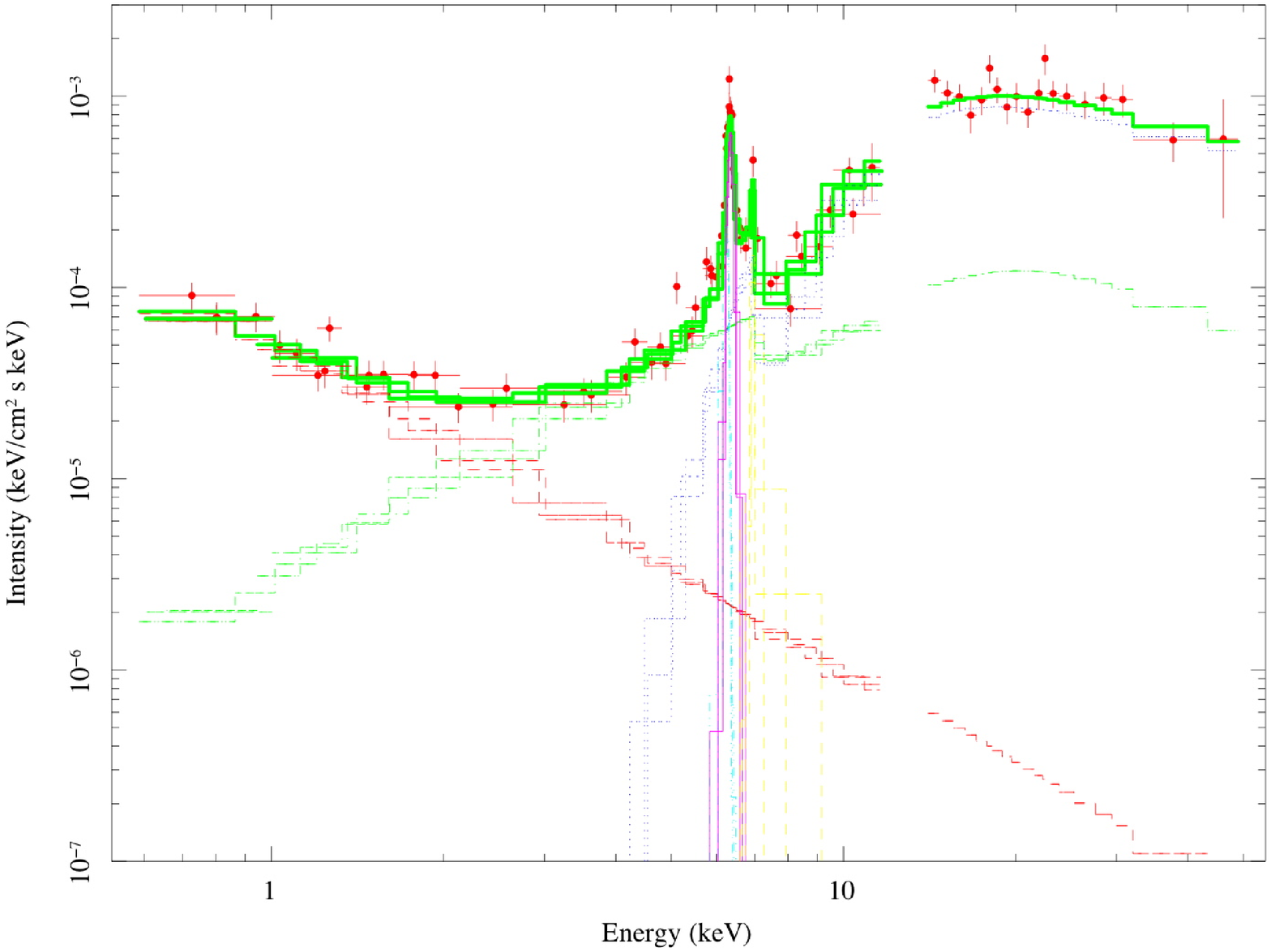}
\end{tabular}
  \end{center}
  \caption{Example of Suzaku X-ray spectra of two buried AGNs.
Note the dominating reflection component (which is curved at high energy)
and the strong iron line. For both objects (SWIFT J0601.9-8636 and NGC 5728
left and right respectively) the scattering efficiency is very low 
(0.2--2\,\%).
Adapted from \cite{ueda07} (left) and \cite{comastri07} (right).
 \label{fig:buried}}
\end{figure*}

However, a major discovery was achieved by INTEGRAL and {\em Swift} and 
this is the detection of ``buried'' super-massive black holes 
\citep{ueda07,comastri07}. These are AGNs for which the reflection component
dominates over the transmitted one. If the ratio of the normalizations of these 
two components is interpreted as the solid angle covered by the reflecting
medium (i.e. R$=\Omega/2\pi$) then this value exceeds 1. This would imply
that  part of the nuclear emission is  blocked by 
nonuniform material  along the line of sight even above 10 keV.
The extremely low scattering efficiency which has been found for
these objects \citep[e.g. 0.5-2\,\%][]{ueda07,comastri07} implies
a torus half-opening angle of $<20^{\circ}$ in contrast to the
classical $30^{\circ}-40^{\circ}$ expected in the framework of
AGN unified model \citep[e.g.][]{antonucci93}.
The large equivalent widths ($\sim$1\,keV, see Fig.~\ref{fig:buried}) 
of the iron K$_{\alpha}$ line
and the absence or weakness of [O III] lines confirm this interpretation.

In order to recognize these objects as such, one needs sensitive and 
broad-band X-ray coverage. Thus it might well be that a few (or many)
of these objects are hiding among the INTEGRAL and {\em Swift}/BAT 
survey sources. According to \cite{winter08} these objects
might be a relevant fraction, $\sim25$\,\%, of the total AGN population
of the local Universe. The question whether this is the {\it missing}
source population (i.e. the one needed to explain the totality
of the CXB emission at its peak) is a difficult one to ask.
The answer might come only from a sample of buried AGN large enough
to derive its luminosity function.
Until that time \cite{dellaceca08b} derived the first luminosity function
of Compton-thick AGNs as a difference between the Optical luminosity
function of AGNs derived by \cite{simpson05}
and their X-ray luminosity function of AGNs. They find that the space
density of Compton-thick objects is twice the density of Compton-thin AGNs
with an hard upper limit to four times this density given by the limit
imposed by the local black hole mass density \citep{marconi04}.

\subsection{Compton-thick AGNs in the high-z Universe}
\label{sec:ir}
\begin{figure*}[ht!]
  \begin{center}
  \begin{tabular}{cc}
    \includegraphics[scale=0.95,clip=true,trim= 240 0 0 00]{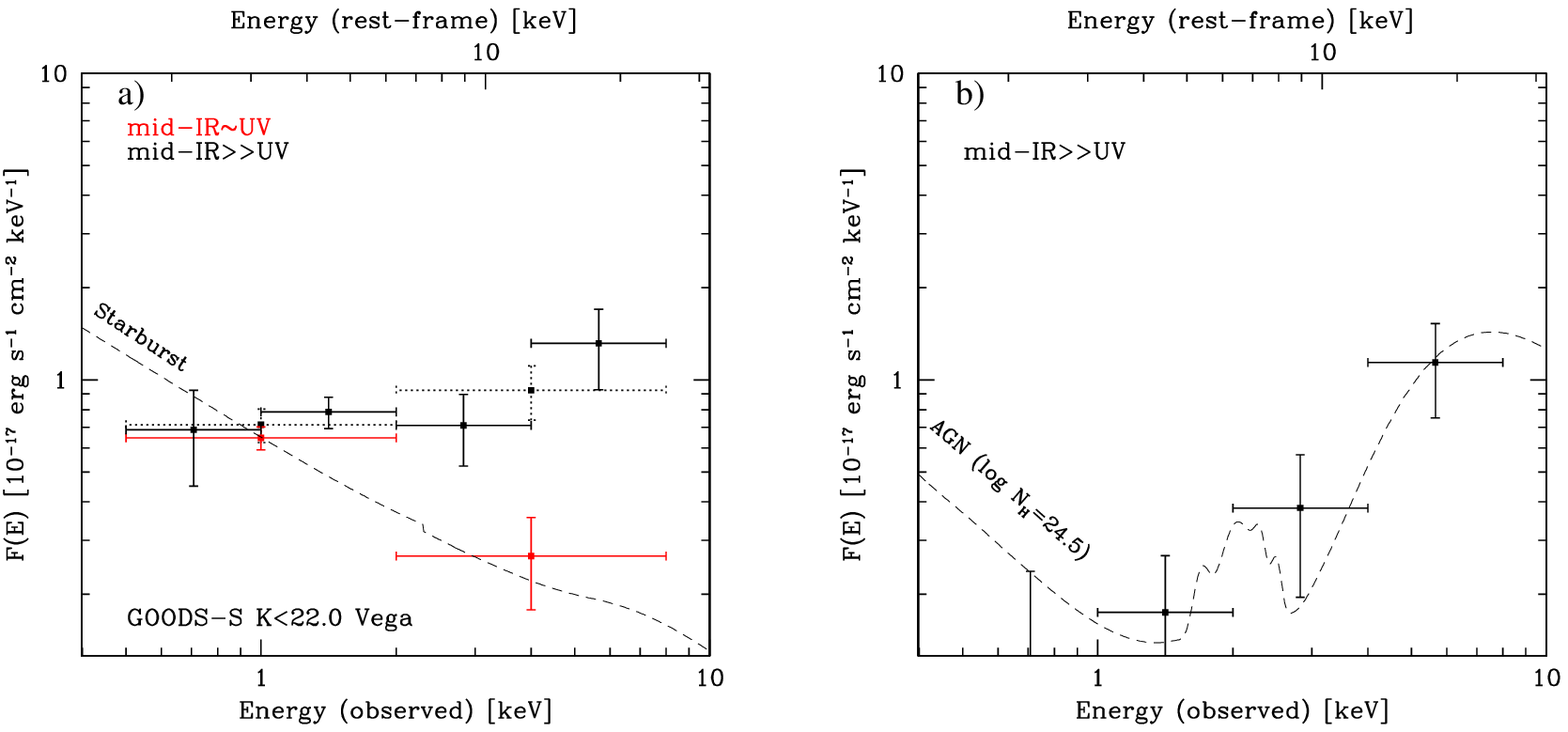} 
  	 \includegraphics[scale=0.40]{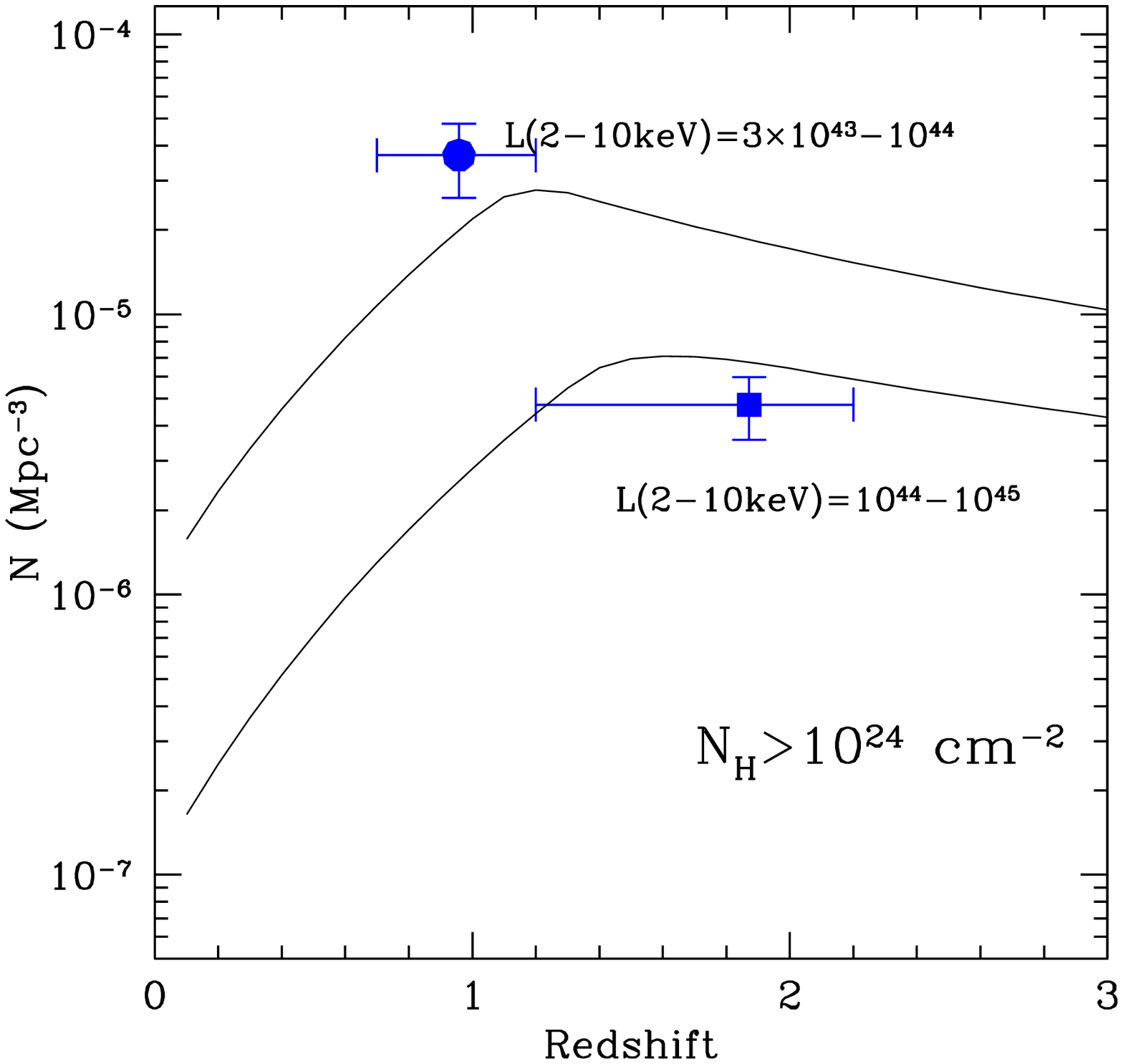}
\end{tabular}
  \end{center}
  \caption{
{\it Left:} Stacked X--ray spectrum of mid-IR excesses detected in the
Chandra deep fields. The spectrum is compatible with the spectrum
of a Compton-thick AGN. Adapted from \cite{daddi07}.
{\it Right:}Volume density of mid-IR selected sources in the COSMOS
field \citep{fiore08b}. The solid lines are predictions from the model of 
\cite{gilli07}.
 \label{fig:midir}}
\end{figure*}

As already seen, selecting Compton-thick objects is extremely
difficult even in hard X-rays. There is however a possibility
to recover obscured AGNs thanks to the reprocessing of the AGN
UV emission in the infrared (IR) band. Thus, selecting bright mid-IR
sources which are faint in the Optical, and thus likely to be obscured,
 might be rewarding \citep[e.g.][]{martinez05,houck05}.
Such window of opportunity has been recently opened by the 
Spitzer telescope. Indeed, using Spitzer \cite{martinez05}
has found that bright mid-IR sources (F(24$\,\mu$m)>0.3\,mJy)
with faint optical and near-IR counterparts are likely to be highly
obscured type 2 QSOs. This approach has been extensively applied to Spitzer
observations of the  Chandra Deep Fields \cite[][]{daddi07,fiore08}. 
In these fields, the X-ray coverage is deep and therefore it allows
to probe the nature of these mid-IR ``excesses''. A few of these
sources have a direct X--ray detection which might indicate that
they are obscured sources\footnotemark{}.\footnotetext{Given the signal-to-noise ratios it is very difficult to say whether these sources are simply
obscured (N$_H>10^{22}$\,cm$^{-2}$) or are Compton-thick 
(N$_H>10^{24}$\,cm$^{-2}$).} The average X-ray properties of a class 
of sources can be studied using the stacking technique (i.e. summing
the signal of different sources). The stacked X--ray spectrum 
(shown in the left panel of Fig.~\ref{fig:midir})
of mid-IR excesses is compatible
with the one of a Compton-thick AGNs \citep[][]{daddi07,fiore08}.
Since most of these sources have either a photometric or a measured
redshift, it becomes possible to estimate their volume density 
(see right panel of Fig.~\ref{fig:midir}).
This, at the average redshift of these samples (z$\sim$2), turns out to be 
of the same order of that of X-ray detected AGNs. This would imply
that the mid-IR selection is a powerful technique to recover
the population of Compton-thick objects which is not detected
even in the deepest X--ray surveys.

However, the complex selection criteria leave some uncertainty about the
true nature of these sources. Indeed, mid-IR excesses might be produced
by powerful sturburst galaxies \cite[e.g.][]{fiore08}. In order to
remove all uncertainties one would need to obtain an IR and an X--ray
spectra for all these sources. Given the large redshift and the number of 
sources this is not always feasible.
At least in the case of HDF-OMD49, a  Spitzer object at z=2.21,
the IR and X--ray spectra show that this source is 
very likely a Compton-thick AGN \cite{alexander08}. However, recently \cite{pope08},
using a compbination of IR bands and mid-IR spectroscopy,
determined that the majority of the dust obscured galaxies\footnotemark{}
\footnotetext{Dust obscured galaxies are selected in such a way that
90\,\% of them would meet the selection criteria of \cite{fiore08}.} 
are dominated by star formation rather than AGN activity.
Thus, while mid-IR selection might represent a powerful tool
to recover Compton-thick AGNs at large redshifts, it might be that
the contamination due to powerful starburst galaxies is at the moment
affecting the estimates of the  space density of highly-obscured AGNs.

\section{Reflection and Cut-off properties of AGNs}

Reflection and high-energy cut-off are with  intrinsic
absorption the main spectral properties of AGNs. Studying them
is, however, not as simple as studying intrinsic absorption.
This is because both reflection and high-energy cut-off produce
a weaker spectral signature and with low signal to noise data
they tend to be degenerate. The optimum solution would be to use high-quality
broad band data which allow to disentangle all three spectral components.
Such approach has been recently 
used for type-1 and type-2 AGNs by \cite{panessa08}
and \cite{derosa08} respectively.
Both groups find evidences for spectral cut-offs which are at energies
$<$300\,keV and thus at odds with previous studies in 
the 1-500\,keV range \citep{zdziarski95,gondek96} which located
the cut-off at $>$500\,keV. In most cases a strong reflection component
and an Iron line are  found, although they appear to be inconsistent 
with the measured absorbing column densities \cite{derosa08}. 
This might indicate that the absorbing and the reflecting material
are not the same or that the absorbing/reflecting material is not
uniform (i.e. a clumpy torus model). 
\cite{dadina08} analyzed a sample of 105 Seyfert objects detected
by BeppoSAX and thus with simultaneous broad-band
coverage in the 2-100\,keV band. He finds a systematic difference
between Seyfert 1 and 2 galaxies with Seyfert 1s having a steeper
spectrum, a larger reflection component and a lower energy spectral
cut-off. This is in agreement with what found for a smaller
sample of {\em Swift}/BAT sources \citep{ajello08a}.
While the different strengths of the
reflection components might be explained by the
different viewing angles of the reflecting material in these
objects, the different spectral cut-offs might signal a break down
of the AGN unified model. Indeed, in thermal Compton models
\citep[e.g.][]{zdziarski95} the cut-off energy is related to the 
thermal energy of the electrons populating the hot corona above the
accretion disk and in this case it would imply that this energy
is lower in Seyfert 1 with respect Seyfert 2.
However, I remark that with low signal to noise spectra is very
difficult to disentangle all the different spectral components.
The best way would be to select a flux-limited sample of bright
Seyfert galaxies for which a 0.1--500\,keV data exist.
While thanks to XMM-Newton, {\it Swift} and INTEGRAL it is possible
to assemble high-quality data up to $\sim200$\,keV, above this energy
there are essentially no data. In this framework the detection of
Seyfert galaxies above 200\,keV \citep[e.g.][]{bouchet08} might
play an important role as it would allow to disentangle the 
spectral cut-off  from the other spectral features.

\section{Unveiling the hard X-ray sky: the SIX survey}
\label{sec:six}
INTEGRAL/IBIS and {\it Swift}/BAT will likely be able to expose
 a significant area of the extragalactic sky
 down
to the 0.5\,mCrab\footnotemark{}
\footnotetext{As a reference 1\,mCrab in the 15--55\,keV band is equivalent
to $\sim$1.27$\times10^{-11}$\,erg $cm^{-2}$ s$^{-1}$.}
 flux level. BAT
will achieve it thanks to its continuos all-sky coverage, while
INTEGRAL will do it through its {\it Key Programme} observations
(e.g. see the 6\,Ms observation of the North Ecliptic Pole).
Simbol-X, {\it NuSTAR} and NeXT \citep[][respectively]{ferrando06,harrison05,takahashi08}, 
all focusing hard X-ray telescopes, will be able to sample fluxes
of 10$^{-13}$--10$^{-14}$\,erg cm$^{-2}$ s$^{-1}$.
Until that time, no other scheduled mission\footnotemark{}\footnotetext{The EXIST mission under study by NASA would probe
the $5\times10^{-13}$\,erg cm$^{-2}$ s$^{-1}$ fluxes \citep{grindlay05}.}
will probe intermediate fluxes.

\begin{figure*}[ht!]
  \begin{center}
  \begin{tabular}{cc}
    \includegraphics[scale=0.41]{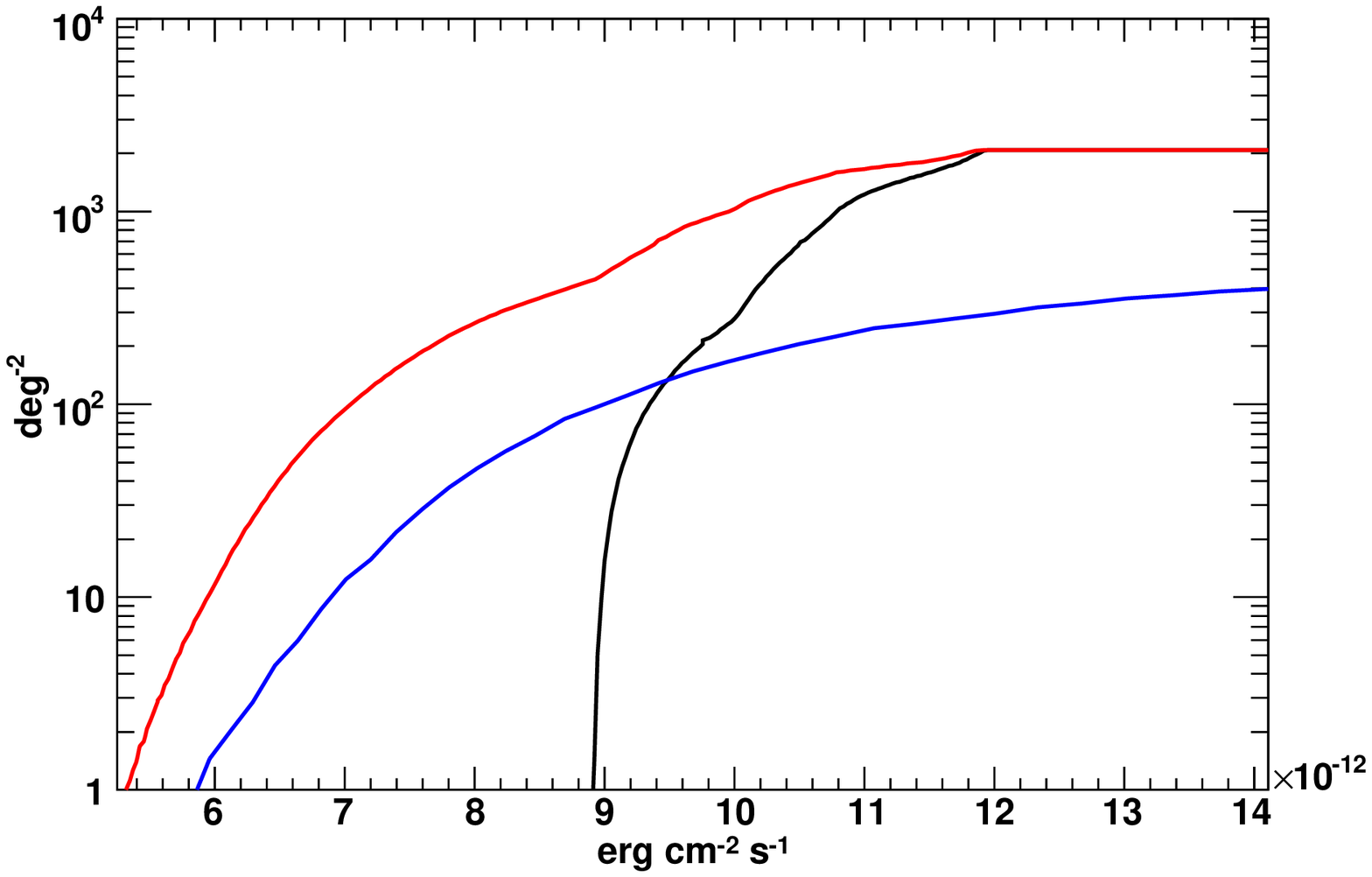} 
  	 \includegraphics[scale=0.41]{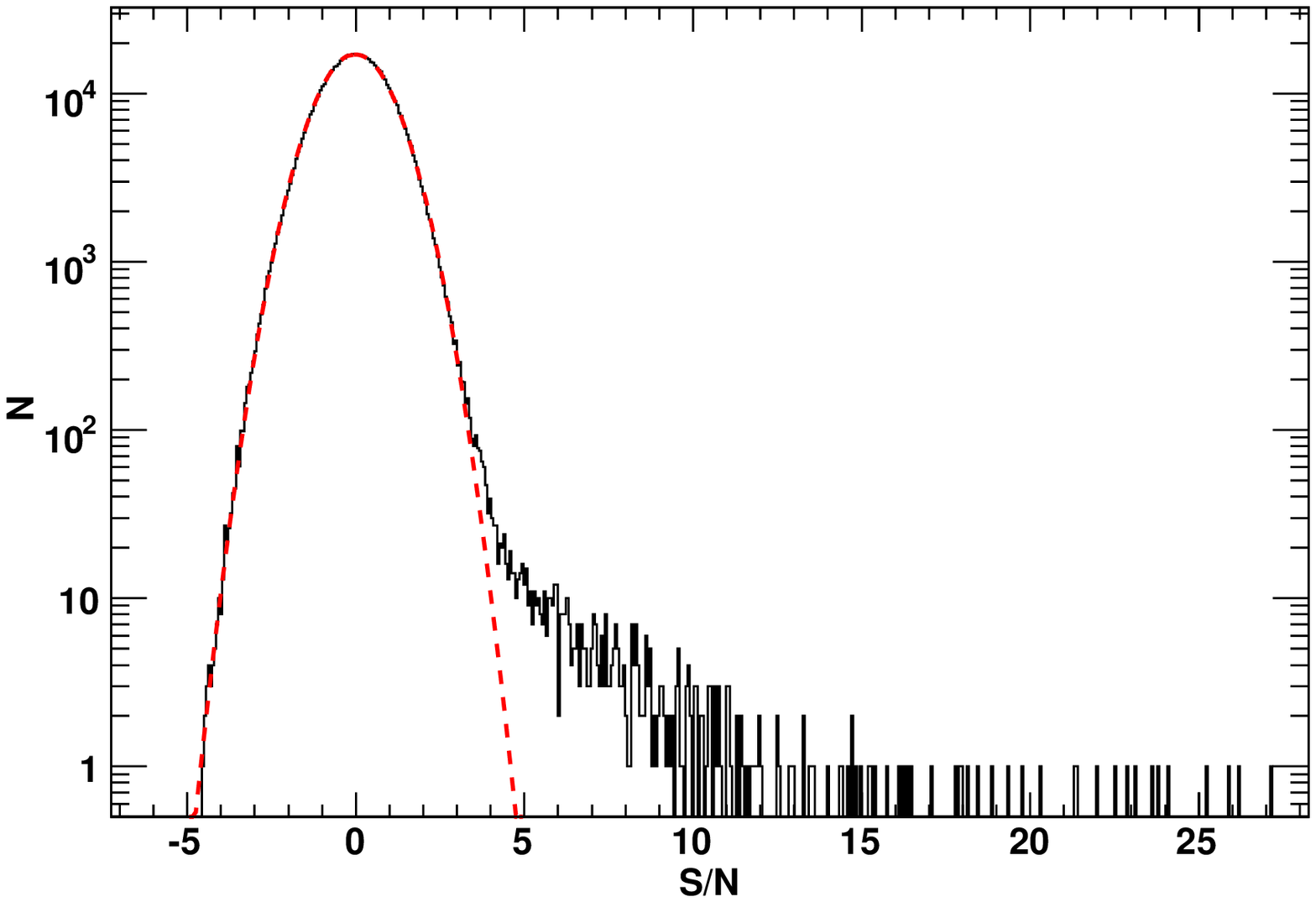}
\end{tabular}
  \end{center}
  \caption{Performances of the SIX survey on a test field. The field chosen
is the North Ecliptic Pole (INTEGRAL Key programme) surveyed by INTEGRAL
to 1.7\,Ms. The left panel shows the BAT (black), IBIS (blue) and
SIX (red) sky coverages. The limiting sensitivity of the SIX mosaic,
in this example, is better than 0.5\,mCrab.
The right panel shows the distribution of pixel significances in the 
SIX-NEP mosaic. The dashed line is an overlaid Gaussian with $\sigma$=1.0.
The long tail at positive significances is given by real sources detected
in the mosaic. 
 \label{fig:six}}
\end{figure*}

There is however a way to sample fluxes weaker than the ones sampled by
the INTEGRAL/IBIS and {\em Swift}/BAT surveys and 
this is represented by the {\em Swift}+INTEGRAL X-ray (SIX) survey
(Ajello et al. in prep.).
The SIX survey is obtained as a natural combination of both surveys.
Indeed, as it is shown in Tab.~\ref{tab:comp}, the performances of both
instruments, BAT and IBIS, are similar in terms of sensitivity
for deep extragalactic exposures. The main difference is represented
by the point spread function (PSF) which is better for INTEGRAL.
However, as long as one avoids crowded regions (i.e. the Galactic plane)
this issue does not matter. The advantage of the SIX survey is not only that it
allows to combine the exposures to obtain a deeper one, but that
it smoothes out the relative systematic errors that both instruments have.
As a test field to demonstrate the potentiality of the SIX survey
we chose the North Ecliptic Pole (NEP) field. This field has been
proposed and approved as an INTEGRAL key-programme because of the
absence of bright sources which could worsen the sensitivity.
At the moment of this writing, the NEP field has been
surveyed to 1.7\,Ms (of the 6\,Ms requested). Using two years
of {\em Swift}/BAT survey data \cite[][]{ajello09} we get an 
exposure on the same region of $\sim4$\,Ms.
Left panel of Fig.~\ref{fig:six} shows the difference between the BAT
and the IBIS surveys. Indeed, while IBIS reaches a deeper sensitivity
at the centre of the mosaic, the BAT survey has shallower, but much
more uniform exposure on the whole area. The SIX survey joins the best
of both worlds yielding already in this simple test a sensitivity 
better than 0.5\,mCrab. As expected by the combination of two
surveys with different systematic errors, the noise is well behaved
and the distribution of signal-to-noise ratios is consistent with
a normal Gaussian distribution (see right panel of Fig.~\ref{fig:six}).
The number of detected sources above 5\,$\sigma$ is 18 which is a
large number considering that it has been detected in an INTEGRAL
mosaic ($\sim$0.5 sr) away from the Galactic plane.
This test shows the potentiality of this approach fully as it has
already produced the most sensitive hard X-ray survey to date.
A much better sensitivity is, reasonably, expected when the NEP
field will be surveyed by INTEGRAL to the planned 6\,Ms.

\begin{table}[ht]
\centering
\caption{Comparison of the BAT and ISGRI instruments. Sensitivities
are real in-flight performances and were derived for BAT 
and ISGRI by Ajello et al. (2008a) and Bassani et al. (2006) respectively.
\label{tab:comp}}
\begin{tabular}{lcc}
\hline
\hline
                     & {\bf BAT}         & {\bf ISGRI} \\
\hline
PSF (arcmin)         &  22               & 12          \\
FOV (deg$^2$)        &  4500             & 400  \\
Energy range (keV)   &  15--200          & 16--300 \\
5\,$\sigma$ Sensitivity  in 1\,Ms (mCrab) &  0.9      & 0.8 \\     
\hline
\end{tabular}
\end{table}

\section{Conclusions}
Thanks to INTEGRAL and {\em Swift} the hard X--ray sky looks  bright.
For the first time in X--ray astronomy, there is an all-sky X-ray selected
sample of AGNs which constitutes an unbiased census of AGNs
in the local Universe. We have learned that contrary to the expectation,
from the unified model, the fraction of absorbed objects, in the local
Universe, is $\sim$50\,\%.
As deep surveys have shown, the ratio of absorbed sources to the total
population is a function of intrinsic luminosity. This has been interpreted
as a decrease of the covering factor of the circomnuclear dust as a 
function of luminosity and finds a natural explanation in the framework
of the clumpy torus model which is, as of today, well established.

Despite their lack, Compton-thick AGNs are still required to explain a
substantial part of the Cosmic X-ray background around its peak.
One of the main discoveries of INTEGRAL and {\em Swift} is to
have unveiled a new population of Compton-thick AGNs. These are AGNs
with an intense reflection component which seems to come from a material
which covers a large solid angle around the central source. The current
estimates show that these 'buried' super-massive black hole
might be a relevant fraction ($\sim$25\,\%) of the total population
of AGNs, although given their low scattering efficiency their contribution
to the CXB spectrum seems small. At the same time, large populations of
Compton-thick objects can be recovered detecting the reprocessing,
by dust, of UV nuclear emission into IR. The current estimates
of their space density at large redshift indicates that Compton-thick
AGNs are at least as numerous as the remainder AGN population.
The number density of detected Compton-thick objects in the local Universe
and in the high-redshift one are in agreement with predictions of
population synthesis models showing that we are close to solve the
mystery of the generation of the Cosmic X-ray Background. If this
view is correct, hard X-ray ($>$15\,keV) focusing telescopes
will detect many Compton-thick AGNs at intermediate redshifts.

\acknowledgments
The processing of INTEGRAL data, in the framework of the SIX 
survey, rests entirely on  E. Bottacini whom
is here warmly acknowledged. Most of the work presented here comes
from my PhD thesis and I acknowledge all members of the gamma group
at MPE  as well as the members of the  BAT-team at Goddard.

\bibliographystyle{pos}
\bibliography{/Users/marcoajello/Work/Papers/BiblioLib/biblio}

\end{document}